\documentclass[12pt]{article}
\usepackage[a4paper,left=1.5748in,right=0.59055in,top=1.5748in]{geometry}
\usepackage{amssymb}
\usepackage{setspace}
\usepackage{amsmath}
\usepackage{enumerate}

\newtheorem{thm}{Theorem}
\newtheorem{prop}[thm]{Proposition}
\newtheorem{lem}[thm]{Lemma}

\newtheorem{exam}{Example}

\newcommand{\pf}{\noindent {\bf Proof.  }}
\newcommand{\qed}{\hfill $\Box$ \\}

\def\0{{\mathbf 0}}
\def\C{{\mathcal C}}

\newcommand{\F}{\mathbb{F}}
\newcommand{\Z}{\mathbb{Z}}

\begin{document}

\title{ On self-dual double negacirculant codes}
\author{Adel Alahmadi\thanks{Math Dept, King Abdulaziz University, Jeddah, Saudi Arabia, {Email: \tt adelnife2@yahoo.com}},
Hatoon Shoaib\thanks{Math Dept, King Abdulaziz University, Jeddah, Saudi Arabia,{Email: \tt hashoaib@kau.edu.sa }},
Patrick Sol\'e\thanks{ CNRS/LTCI, Telecom ParisTech, Universit\'e de Paris-Saclay, 75 013 Paris, France \& Math. Dept, King Abdulaziz University, Jeddah, Saudi Arabia.
{Email: \tt sole@enst.fr}}
}
\date{}
\maketitle
\begin{abstract} Double negacirculant  (DN) codes are the analogues in odd characteristic of double circulant codes.
Self-dual DN codes of odd dimension are shown to be consta-dihedral. Exact counting formulae are derived for DN codes. 
The special class of length a power of two is studied by means of Dickson polynomials, and is shown to 
contain families of codes with relative distances satisfying a modified Gilbert-Varshamov bound.
 
\end{abstract}

\vspace*{1cm}

\bf Key Words\rm : quasi-twisted codes, dihedral group, Dickson polynomials


\section{Introduction}
A matrix $A$ over a finite field $\F_q$ is said to be {\em negacirculant} if its rows are obtained by successive negashifts from the first row.
A {\em negashift} maps the vector $(x_0,\dots,x_{n-1})\in \F_q^n$ to $(-x_{n-1},x_0,\dots,x_{n-2}).$

In this paper we consider {\em double negacirculant} (DN) codes over finite fields, that is $[2n,n]$ codes with generator matrices of the shape $(I,A)$ with
$I$ the identity matrix of size $n$ and $A$ a negacirculant matrix of order $n.$
For instance, the famous {\it tetracode}, a self-dual MDS ternary code of length $4$ \cite{HP} is DN.
This construction was introduced in \cite{HG} under the name {\em quasi-twisted code}. We prefer to reserve this term for the more general class of codes described in \cite{Y}.
A code of length $N$ is quasi-twisted of index $\ell$ for $\ell \mid N,$ and co-index $m=\frac{N}{\ell}$
if it is invariant under the power $T_{\alpha}^\ell$ of the {\em constashift} $T_{\alpha}$ defined as

$$T_{\alpha}: (x_0,\dots,x_{N-1}) \mapsto (\alpha x_{N-1},x_0,\dots,x_{N-2}).$$
Such a code affords a natural structure of module over the auxilliary ring $$R(m,\F_q)=\frac{\F_q[x]}{(x^m-{\alpha})}.$$ In other words, it can be regarded as a code of length $\ell$ over the ring
$R(m,\F_q).$ An algebraic way to study such a code is to decompose the semilocal ring $R(m,\F_q)$ as a direct sum of local rings by the Chinese Remainder Theorem \cite{Y}, thus following the approach initiated for quasi-cyclic codes
in \cite{LS}. The benefit of this technique is to reduce the study of QT codes to that of shorter codes over larger alphabets. Besides, the study of duality is made transparent, 
thus allowing the construction of self-dual QT codes, as in, for instance, \cite{Y}. The number of rings occuring in the decomposition of $R(m,\F_q)$ equals the number of irreducible factors of
$x^m-a.$
In the following section we focus on self-dual DN codes. These have been explored numerically in \cite{GHM,HHKK}. First, we show that such codes are consta-dihedral, in the
sense that they are invariant under a monomial representation of the dihedral group of order $2n.$ This definition is related to but different from that of \cite{SR}. Next, we specialize
further by assuming that $n$ is a power of two.
For some specific alphabets, it can be shown that in that case $x^n+1$ can be factored into a product of two irreducible polynomials \cite{M,LN}. This is a favorable situation to
apply the Chinese Remainder Theorem approach of \cite{LS,Y}, as the decomposition of $R(m,\F_q)$ contains only two terms. It allows to derive exact enumeration formulae and, from there, using the so-called 
expurgated random coding technique, to
give an asymptotic lower bound on the minimum distance of these DN codes. This is an analogue of the Varshamov Gilbert bound. Other similar asymptotic estimates use non
constructive lengths and factorization of $x^n+1$ \cite{C,BM}. The analogous result on double circulant codes in even characteristic relies on Artin conjecture \cite{CPW}, or an
intricate induction \cite{K}.

The material is organized as follows. The next section collects the necessary notions and notations. Section 3 studies the automorphism group of these codes. Section 4
derives the factorization  of $x^n+1$ over $\F_q$ when $n$ is a power of $2$ and $q$ is odd. Section 5 contains enumeration formulae and asymptotics. Section 6 concludes the article
and points the way to some challenging open problems.
\section{Definitions and notation}
\subsection{Codes}
Let $\F_q$ denote the finite field of order $q.$ We assume throughout that $q$ is odd. In the following, we will consider codes over $\F_q$ of length $2n$
with $n$ odd and coprime to $q$. Their generator matrix $G$ will be of the form $G=(I,A)$ where $I$ is the identity matrix of
order $n$ and $A$ is an $(n\times n)$-negacirculant matrix. We call such codes {\em double negacirculant} (DN) codes. We will denote by $\C_a$ the DN code
with first row of $A$ being the $x-$expansion of $a$ in the ring $R(n,\F_q).$
\\ \ \

If $C(n)$ is a family of codes of parameters $[n,k_n,d_n],$ the {\em rate} $R$ and {\em relative distance} $\delta$ are defined as
$$R=\limsup_{n \rightarrow \infty}\frac{k_n}{n},$$
and
$$\delta=\liminf_{n \rightarrow \infty}\frac{d_n}{n}.$$
Both limits are finite as limits of bounded quantities.
Such a family of codes is said to be {\it good } if $R\delta \neq 0.$

Recall the $q-$ary {\em entropy function} defined for $0<x< \frac{q-1}{q}$ by $$ H_q(x)=x\log_q(q-1)-x\log_q(x)-(1-x)\log_q(1-x).$$
This quantity is instrumental in the estimation of the volume of high-dimensional Hamming balls when the base field is $\F_q.$
The result we are using is that the volume of the Hamming ball of radius $xn$ is, up to subexponential terms,  $q^{nH_q(x)},$ when $0<x<1$ and $n$ goes to infinity
\cite[Lemma 2.10.3]{HP}.
\subsection{Groups}
The \textit{symmetric} group $S_n$ is the group of permutations of $n$ objects.
The \textit{dihedral} group ${\cal D}_n,$ is defined as the unique group of order $2n$ on two generators $r$ and $s$ of respective orders
$n$ and $2$ and satisfying the relation $srs=r^{-1}.$ A code of length $2n$ is called \textit{dihedral} if it is invariant under ${\cal D}_n$
acting transitively on it's coordinate places.

Recall that a {\em monomial} matrix over $\F_q$ of order $g$ has exactly one nonzero element per row and per column. 
The monomial matrices form a group $M(g,q)$ of order $g!(q-1)^g$ under multiplication.  This group is abstractly isomorphic to the wreath product $\Z_{q-1}\wr S_g.$

By a {\em monomial representation} of a group $G$ over $\F_q$ we shall mean a group morphism from $G$ into $M(g,q).$
A code of length $2n$ will be said to be {\em consta-dihedral} if it is held invariant under right multiplication by a monomial representation of the dihedral group ${\cal D}_n.$
An alternative, but related definition can be found in \cite{SR}.
\subsection{Polynomials}
The {\em Dickson polynomial} (of the first kind)  are given by $D_0(x,\alpha) = 2$, and for $n > 0,$ by 

  $$  D_n(x,\alpha)=\sum_{p=0}^{\lfloor n/2\rfloor}\frac{n}{n-p} \binom{n-p}{p} (-\alpha)^p x^{n-2p}. $$
  The $D_n$ satisfy the identity

   $$ D_n(u + \alpha/u,\alpha) = u^n + (\alpha/u)^n .$$
\section{Symmetry}

Let $M_n(q)$ denote the set of all $n \times n$ matrices over $\F_q.$\\
\begin{lem}

Let $A$ be an $(n\times n)-$negacirculant matrix over $\F_q.$ Then there exists an $(n \times n)-$generalized permutation matrix $P$ such that
$PAP=A^T.$
\end{lem}

\pf
Consider the following permutation is $S_n$:\\
\centerline{$\pi =(2,n)(3,n-1)...(\frac{n-1}{2},\frac{n+5}{2})(\frac{n+1}{2},\frac{n+3}{2}).$}\\
Then the generalized permutation matrix $P_{\pi}$ corresponding to the permutation $\pi$, is\\
$$P_{\pi}= \left(
            \begin{array}{c}
            e_{\pi(1)} \\
            e_{\pi(2)} \\
            . \\
            . \\
            . \\
            e_{\pi(n)} \\
            \end{array}
          \right)
          =\left(
             \begin{array}{c}
             e_1 \\
             e_n \\
             . \\
             . \\
             . \\
             e_2 \\
             \end{array}
           \right)
           =\left(
              \begin{array}{ccccc}
                1 & 0 & ... & 0 & 0 \\
                0 & 0 & ... & 0 & \alpha \\
                0 & 0 & ... & \alpha & 0 \\
                . & . & ... & . & . \\
                . & . & ... & . & . \\
                . & . & ... & . & . \\
                0 & \alpha & ... & 0 & 0 \\
              \end{array}
            \right)
            =P
$$
where $\alpha^2=1.$ Then it easy to observe that,
$PAP=A^T.$

\qed

\begin{thm}
 For $n \geq 3,$ and $q$ odd, every self-dual quasi twisted code $\C$ of length $2n$ over $\F_q$ is consta-dihedral.
\end{thm}

\pf
Let $C$ be a self-dual double negacirculant code of length $2n$ with generator matrix $G=(I,A)$ with $A$ negacirculant and $AA^T=-I$. 
Computing $A^TG=(A^T,-I)$ and conjugating by $P$ of Lemma 1 we get $PA^TGP=(A,-I)$. 
Define the antiswap involution $s$ by the rule $s(x,y)=s(y,-x)$, where $x,y$ are vectors of length $n$ over $\F_q.$ Note that $s^2=1$.
Clearly $ s \in M(2n,q).$ 
Thus $\pi s \in M(2n,q)$ and it preserves $C.$ A monomial representation of ${\cal D}_n$ is then $\langle \tau , \pi s\rangle.$ Thus $C$ is consta-dihedral.
\qed
\section{Factorizations}
The complete factorization of $x^{2^n}+1$ over $\F_q$ with $q \equiv 3$ (mod 4) is given in the following theorem \cite{M}.
\begin{thm}\label{1}
Let $q \equiv 3 \pmod{4},$ where $q=2^Am-1$, $A \geq 2$, $m$ is odd integer. Let $n \geq 2$,\\
(a) If $n < A$, then $x^{2^n}+1$ is the product of $2^{n-1}$ irreducible trinomials over $\F_q$
$$x^{2^n}+1= \prod_{\gamma \in \Gamma}(x^2+ \gamma x +1),$$
where $\Gamma$ is the set of all roots of $D_{2^{n-1}}(x,1)$.\\
(b) If $n \geq A$, then $x^{2^n}+1$ is the product of $2^{A-1}$ irreducible trinomials over $\F_q$
$$ x^{2^n}+1 = \prod_{\delta \in \Delta}(x^{n-A+1}+ \delta x^{n-A}-1),$$
where $\Delta$ is the set of all roots of $D_{2^{A-1}}(x,-1)$ in $\F_q.$
\end{thm}

\begin{exam}

If $q=3$ i.e. $q \equiv 3 \pmod{4},$ then $q=2^{2}.1-1$ implies that $A=2$, $m=1$, and $D_{2}(x,-1)=\{1,2\}$
, then by Theorem \ref{1}:
$$x^{2^{n}}+1=(x^{2^{n-1}}+x^{2^{n-2}}+2)(x^{2^{n-1}}+2x^{2^{n-2}}+2)$$
\end{exam}

We need the analogous factorization theorem when $q \equiv 1$ (mod 4). We prepare for the proof by an arithmetic Lemma. If $r,s$ are two integers, we denote by $ord_r(s)$
the smallest integer $i$ such that $s^i \equiv 1 \pmod{r}.$

\begin{lem}\label{above}
 If $q=2^{A+1}m+1$, with $m,$ odd and $A$ integer, then $ord_{2^{n+1}}(q)=2^{n-A}.$
\end{lem}

\pf
We derive
\begin{equation}\label{fact}
q^{2^{n-A}} \equiv 1 \pmod{2^{n+1}}
\end{equation}
by induction on $n$.\\
For $n=A+1$, note that, by definition, $q \equiv 1 \pmod{2^{A+1}},$ implying $q^2 \equiv 1 \pmod{2^{A+2}}.$ Therefore (\ref{fact}) is true for $n=A+1$.\\
Assume that (\ref{fact}) is true for $n=k+A+1$, then we have
$q^{2^{k+1}}= 1+ \beta 2^{A+k+2},$ for some integer $\beta,$
and squaring, we get $q^{2^{k+2}}= 1+ 2\beta 2^{A+k+2}+\beta^2 2^{A+k+2},$ which implies (\ref{fact}) for $n=k+A+2.$
This shows that $ord_{2^{n+1}}(q)$ is a power of two dividing $2^{n-A}.$
Using a similar induction on $n$ it can be shown that 
$$q^{2^{n-A-1}} \equiv 1+2^n\pmod{2^{n+1}}.$$
Hence $ord_{2^{n+1}}(q)=2^{n-A}.$
\qed

\begin{thm}\label{2}
 Let $q \equiv 1 \pmod{4},$ where $q=2^Am+1$, $A \geq 2$, $m$ is odd integer. Denote by $U_k$ is the set of all primitive $2^k$th roots of unity in $\F_q.$ If $n \geq 2$, then\\
 \begin{itemize}
  \item[(a)] If $n \leq A$, then $ord_{2^{n+1}}(q)=1$ and $x^{2^n}+1$ is the product of $2^n$ linear factors over $\F_q$
$$x^{2^n}+1 = \prod_{u \in U_{n+1}}(x+u).$$
  \item[(b)] If $n \geq A+1$, then $ord_{2^{n+1}}(q)=2^{n-A}$ and $x^{2^n}+1$ is the product of $2^A$ irreducible binomials over $\F_q$ of degree $2^{n-A}$
$$ x^{2^n}+1= \prod_{u \in U_{A+1}}(x^{2^{n-A}}+ u).$$

 \end{itemize}

\end{thm}

\pf
(a) In this case, $q\equiv 1 \pmod{2^{n+1}},$ we can apply \cite[theorem 2.47]{LN}, with $d=1.$
(b) If $n \geq A+1$, then to apply \cite[theorem 2.47]{LN}, knowing that $\phi(2^{n+1})=2^n$ and $d | 2^n$, we use Lemma \ref{above} to prove that  $ord_{2^{n+1}}(q)=2^{n-A}$
  so that we can take  $d=2^{n-A},$ and therefore that $x^{2^n}+1$ is the product of $2^A$ irreducible polynomials over $\F_q$ of degree $2^{n-A}$.

Writing this identity for $n=A$ by (a) where the factors are linear and substituting $x=x^{2^{n-A}},$ we obtain a factorization of $2^A$ binomials over $\F_q,$ each of degree $2^{n-A}.$
They can be shown to be irreducible by application of \cite[th. 3.75]{LN}. The result follows.
\qed
\begin{exam}
If $q=5$ i.e. $q \equiv 1$ (mod 4), then $q=2^{2}.1+1$ implies that $A=1$, $m=1$, and $U_2=\{2,3\}$
, then by Theorem \ref{2}:
$$x^{2^{n}}+1=(x^{2^{n-1}}+2)(x^{2^{n-1}}+3).$$
\end{exam}

\section{Asymptotics}
\subsection{Enumeration}
The following result is not needed in its full generality for the asymptotic application, but is of interest in its own right. It can be viewed as enumerating negacirculant
matrices of order $n.$

\begin{prop}\label{enum}
 Let $n$ be an integer, and $q$ a prime power coprime with $n$. Suppose, if $n$ is odd, that $-1$ is a square in $\F_q$. Assume that the factorization of $x^n+1$ into irreducible polynomials over $\F_q$ is of the form
$$x^n+1=\alpha \prod_{j=1}^s g_j(x) \prod_{j=1}^t h_j(x)h^*_j(x),$$
with $\alpha \in \F_q$, $n=s+2t$ and $g_j$ a self-reciprocal polynomial of degree $2d_j$, the polynomial $h_j$ is of degree $e_j$ and $*$ denotes reciprocation.
If $n$ is odd, then let $g_1=x+1$. The number of self-dual double negacirculant codes over $\F_q$ is then
$$2\prod_{j=2}^s (1+q^{d_j}) \prod_{j=1}^t (q^{e_j}-1)$$
if $n$ is odd
$$\prod_{j=1}^s (1+q^{d_j}) \prod_{j=1}^t (q^{e_j}-1)$$
if $n$ is even.
\end{prop}

\pf (sketch) We use the Chinese Remainder Theorem decomposition of $R(n,\F_q),$ as explained in the Introduction.
If $n$ is odd and $\F_q$ contains a square root of $-1,$ say $\omega,$ then the factor $g_1=x+1$ of $x^n+1$ yields a term $\F_q$ in that decomposition.
There are two self-dual codes of length $2$ over $\F_q$ that is $<[1,\omega]>,$ and $<[1,-\omega]>.$ More generally, a factor $g_i$ of degree $2d_i$ leads to counting
self dual hermitian codes of length $2$ over $\F_{q^{2d_i}},$ that is to count the solutions of the equation $1+x^{1+q^{d_i}}=0$ over that field. Using the existence
of a root of order $q^{2d_i}-1,$ it can be seen that this equation has $1+q^{d_i}$ solutions. Alternatively, one could specialize the formulae in \cite{RS}.

 In the case of reciprocal pairs $(h_j,h_j^*)$, note that the number of linear codes of length $2$ over some
$\F_Q$ admitting, along with their duals, a systematic form is $Q-1,$ all of dimension $1.$ Indeed their generator matrix is of the form $[1,u]$ with $u$ nonzero.
We conclude by letting $Q=q^{e_j}.$
\\
\qed

\subsection{Distance bounds}

In this section, we assume that $q$ is such that $x^n+1,$ for $n$ a power of $2,$ has only two irreducible factors, say $h'$ and $h''$, and that they are reciprocal of each other.
For convenience, let $K'=\frac{\F_q[x]}{(h')}$ and $K''=\frac{\F_q[x]}{(h'')}.$ These two fields are both isomorphic to $\F_{q^{n/2}}.$
By Theorems 3 and 5, this is the case if $q=4m\pm 1,$ with $m$ odd. For instance this happens if $q=3,5$ but not if $q=7.$
\begin{lem}\label{cover}
 If $u \neq 0$ has Hamming weight $< n$, there are at most $q^{\frac{n}{2}}$ polynomials $a$ such that $u \in \C_a=<[1,a]>,$
 and at most one polynomial $a$ such that $u \in \C_a$ and $\C_a$ is self dual.

\end{lem}

\pf
Let $\C_a=<[1,a]>$, and let $u=(c,d),$ with $c,d$ of length $n$. The condition $u \in \C_a$ is equivalent to the equations,
$$d'=a'c'  \hskip8mm \textrm{over} \hskip3mm K'$$
$$d''=a''c''  \hskip8mm \textrm{over} \hskip3mm K''$$
Then we have:
\begin{itemize}
 \item[(I)]If $\C_a$ is self dual, then $<[1,a']>=<[1,a'']>^{\perp}$ implies $1+a'a''=0$\\
If $d'=c'=0$ ,then $a'$ is undetermined, and if $c'',d'' \neq 0$ ,then $a''=\frac{d''}{c''}$ that implies $a'$ determined and $a'= - \frac{1}{a''}.$
Hence, for given $u=(c,d)$, there is at most one choice for $a$.
\item[(II)] If $\C_a$ is not necessarily self dual then, we have two cases:
\begin{itemize}

\item[(i)]If $c' \neq 0$, then $a' = \frac{d' }{c' }$ has a unique solution.\\
\item[(ii)]If $c' = 0$, then

(a)If $d' \neq 0$, then we have no solution.

(b)If $d' = 0$, then $a'$ is undetermined i.e. we have $q^{\frac{n}{2}}$ choices for $a'$. 
\end{itemize}
Similarly, we have the same solutions for $a''$:
\begin{itemize}
\item[(i)] If $c'' \neq 0$, then $a'' = \frac{d'' }{c'' }$ has a unique solution.\\
\item[(ii)] If $c'' = 0$,  then

(a)If $d'' \neq 0$, then we have no solution.

(b)If $d'' = 0$, then $a''$ is undetermined i.e. we have $q^{\frac{n}{2}}$ choices for $a''$.
\end{itemize}
Hence, for given $u=(c,d),$ there is at most $q^{\frac{n}{2}}$ choices for $a$.

\end{itemize}

\qed

We are now ready for the main result of this paper.

\begin{thm}
 If $q$ is odd integer, and $n$ is a power of $2$, then there are infinite families of:\\
(i) double negacirculant codes of relative distance $\delta$ satisfying  $H_q(\delta)\geq \frac{1}{4}$.\\
(ii) self dual double negacirculant codes of relative distance $\delta$ satisfying $H_q(\delta)\geq \frac{1}{4}$.
\end{thm}

\pf
(i)The double negacirculant codes containing a vector of weight $d\sim \delta n$ or less are by standard entropic estimates and  Lemma \ref{cover} of the order 
$q^{\frac{n}{2}} \times q^{2n H_q(\delta)}$, up to subexponential terms. This number will be less that the total number of double negacirculant codes which is by 
Proposition \ref{enum}  of the order of $q^{n}$.\\
(ii) The double negacirculant codes containing a vector of weight $d\sim \delta n$ or less are by standard entropic estimates 
and Lemma \ref{cover} of the order $1 \times q^{2n H_q(\delta)}$, up to subexponential terms. This number will be less that the total number of self dual double negacirculant codes 
which is by Proposition \ref{enum} of the order of $q^{\frac{n}{2}}$.
\qed

\section{Conclusion}
In this paper, we have considered double negacirculant codes. As noted in \cite{GHM}, they are an alternative to double circulant (DC) self-dual codes over some fields where DC do not exist.
By an old result of \cite{VR}, a DC self-dual code does not exist over fields like $\F_3$ where $-1$ is not a square. This paper can thus be considered as a companion paper of 
\cite{AOS}. The main difference between the two papers is that a factorization of $x^n+1$ into two irreducible polynomials is easier and more elementary to find than a factorization of $x^n-1$ into two irreducible polynomials
which requires $n$ to be a prime for which $2$ is primitive. The existence of infinitely many such $n'$s requires the truth of Artin's conjecture \cite{M}. Thus the present paper
is more explicit and more elementary than \cite{AOS}.

Many questions remain open. The DN codes can be construed as QT codes of index $2.$ In view of the constructions of \cite{HHKK}, it would be of interest to 
consider quasi-twisted codes of index $4.$ At a technical level, it would be interesting to generalize Lemma \ref{cover} to fields $\F_q$ with $q$'s like $q=7,$ where the factorization of $x^n+1$ for $n$ a power of $2,$ 
contains four polynomials or more. More generally, in view of the existence of self-dual quasi-twisted codes for the hermitian product \cite{SJU}, 
generalizing our asymptotics result 
to that setting is a possibility.

\end{document}